\documentstyle[12pt]{article}

\def\hybrid{\topmargin -20pt	\oddsidemargin 0pt
	\headheight 0pt	\headsep 0pt
	\textwidth 6.25in	
	\textheight 9.5in	
	\marginparwidth .875in
	\parskip 5pt plus 1pt	\jot = 1.5ex}

\hybrid

\def\baselinestretch{1.2}

\catcode`\@=11

\def\marginnote#1{}
\newcount\hour
\newcount\minute
\newtoks\amorpm
\hour=\time\divide\hour by60
\minute=\time{\multiply\hour by60 \global\advance\minute by-\hour}
\edef\standardtime{{\ifnum\hour<12 \global\amorpm={am}%
	\else\global\amorpm={pm}\advance\hour by-12 \fi
	\ifnum\hour=0 \hour=12 \fi
	\number\hour:\ifnum\minute<10 0\fi\number\minute\the\amorpm}}
\edef\militarytime{\number\hour:\ifnum\minute<10 0\fi\number\minute}

\def\draftlabel#1{{\@bsphack\if@filesw {\let\thepage\relax
   \xdef\@gtempa{\write\@auxout{\string
      \newlabel{#1}{{\@currentlabel}{\thepage}}}}}\@gtempa
   \if@nobreak \ifvmode\nobreak\fi\fi\fi\@esphack}
	\gdef\@eqnlabel{#1}}
\def\@eqnlabel{}
\def\@vacuum{}
\def\draftmarginnote#1{\marginpar{\raggedright\scriptsize\tt#1}}

\def\draft{\oddsidemargin -.5truein
	\def\@oddfoot{\sl preliminary draft \hfil
	\rm\thepage\hfil\sl\today\quad\militarytime}
	\let\@evenfoot\@oddfoot	\overfullrule 3pt
	\let\label=\draftlabel
	\let\marginnote=\draftmarginnote
   \def\@eqnnum{(\theequation)\rlap{\kern\marginparsep\tt\@eqnlabel}%
\global\let\@eqnlabel\@vacuum}  }

\def\preprint{\twocolumn\sloppy\flushbottom\parindent 2em
	\leftmargini 2em\leftmarginv .5em\leftmarginvi .5em
	\oddsidemargin -.5in	\evensidemargin -.5in
	\columnsep .4in	\footheight 0pt
	\textwidth 10.in	\topmargin  -.4in
	\headheight 12pt \topskip .4in
	\textheight 6.9in \footskip 0pt
	\def\@oddhead{\thepage\hfil\addtocounter{page}{1}\thepage}
	\let\@evenhead\@oddhead	\def\@oddfoot{}	\def\@evenfoot{} }

\def\numberbysection{\@addtoreset{equation}{section}
	\def\theequation{\thesection.\arabic{equation}}}

\def\underline#1{\relax\ifmmode\@@underline#1\else
	$\@@underline{\hbox{#1}}$\relax\fi}

\def\titlepage{\@restonecolfalse\if@twocolumn\@restonecoltrue\onecolum
n
     \else \newpage \fi \thispagestyle{empty}\c@page\z@
	\def\thefootnote{\fnsymbol{footnote}} }

\def\endtitlepage{\if@restonecol\twocolumn \else \newpage \fi
	\def\thefootnote{\arabic{footnote}}
	\setcounter{footnote}{0}}  

\catcode`@=12
\relax

\def\figcap{\section*{Figure Captions\markboth
	{FIGURECAPTIONS}{FIGURECAPTIONS}}\list
	{Figure \arabic{enumi}:\hfill}{\settowidth\labelwidth{Figure
999:}
	\leftmargin\labelwidth
	\advance\leftmargin\labelsep\usecounter{enumi}}}
 \relax
\def\tablecap{\section*{Table Captions\markboth
	{TABLECAPTIONS}{TABLECAPTIONS}}\list
	{Table \arabic{enumi}:\hfill}{\settowidth\labelwidth{Table
999:}
	\leftmargin\labelwidth
	\advance\leftmargin\labelsep\usecounter{enumi}}}
 \relax
\def\reflist{\section*{References\markboth
	{REFLIST}{REFLIST}}\list
	{[\arabic{enumi}]\hfill}{\settowidth\labelwidth{[999]}
	\leftmargin\labelwidth
	\advance\leftmargin\labelsep\usecounter{enumi}}}
 \relax
\makeatletter
\newcounter{pubctr}
\def\publist{\@ifnextchar[{\@publist}{\@@publist}}
\def\@publist[#1]{\list
	{[\arabic{pubctr}]\hfill}{\settowidth\labelwidth{[999]}
	\leftmargin\labelwidth
	\advance\leftmargin\labelsep
	\@nmbrlisttrue\def\@listctr{pubctr}
	\setcounter{pubctr}{#1}\addtocounter{pubctr}{-1}}}
\def\@@publist{\list
	{[\arabic{pubctr}]\hfill}{\settowidth\labelwidth{[999]}
	\leftmargin\labelwidth
	\advance\leftmargin\labelsep
	\@nmbrlisttrue\def\@listctr{pubctr}}}
 \relax
\makeatother
\newskip\humongous \humongous=0pt plus 1000pt minus 1000pt

\newif\ifdtup

\relax

\def\thefootnote{\fnsymbol{footnote}}
\def\iit{\elevenit}
\def\bff{\elevenbf}
\def\s{\sigma}

\def\p{\partial}

\def\a{\alpha}

\def\g{\gamma}

\def\iit{\it}
\def\bff{\bf}
\begin{document}
\renewcommand{\theequation}{\thesection.\arabic{equation}}
\begin{titlepage}
\begin{center}

\hfill CERN-TH.7012/93\\
\hfill hep-th/9309101\\
\vskip .4in

{\large \bf The Dual Faces of String Theory\footnote{Talk presented
at the "Strings 93" Conference, May 1993, Berkeley. To appear in the
Proceedings.}}
\vskip .5in

{\bf Elias Kiritsis}
\vskip .1in
{\em CERN, CH-1211\\
 Geneva 23, SWITZERLAND}
\vskip .3in
\end{center}

\vskip .7in

\begin{center} {\bf ABSTRACT } \end{center}
\begin{quotation}\noindent
Duality symmetries for strings moving in non-trivial
spacetime
backgrounds are analysed. It is shown that, for backgrounds generated
from
compact WZW and coset models, such duality symmetries are exact to
all
orders in
string perturbation theory. A global treatment of duality symmetries
is given, by associating them to the known symmetries of affine
current algebras (
affine-Weyl group and external automorphisms).
It is argued that self-duality symmetries of WZW and coset models
generate the duality symmetries of their moduli space.
Some remarks are presented, concerning the survival of such
symmetries in the non-compact case.
The implications of duality symmetries for string dynamics in
non-trivial/singular  spacetimes are discussed.
\end{quotation}
\vskip 1.0cm
CERN-TH.7012/93 \\
September 1993\\
\end{titlepage}
\vfill
\eject
\def\baselinestretch{1.2}
\baselineskip 16 pt
\noindent
\section{Introduction}
\setcounter{equation}{0}

Strings, being extended objects, sense the target space into which
they are
embedded, in a different way than point particles.
In a compact space this difference appears because, strings,
except from their local excitations that
mimic point particle behavior (``momentum" modes), have ``winding"
excitations where the string wraps around non-contractible cycles of
the
manifold.
The masses of momentum modes  are inversely proportional to the
volume of the
manifold, whereas those of the winding modes are proportional to the
volume,
since it costs energy in order to stretch the string.
Moreover, the string contains oscillating modes that respond to
background fields differently than the center of mass of the string.
In certain cases, the physics of string propagation remains invariant
under a reorganization of the one string Hilbert space and a specific
change in the background. This symmetry is known as duality.
In the simplest possible example, that of a string moving on a
circle,
it was observed that the spectrum of the theory with radius $R$ and
that with
radius $1/R$ are identical, once we interchange winding and momentum
modes, \cite{duality}.

It turns out that such duality symmetries exist (semi-classically)
for
all backgrounds  with isometries, \cite{buscher}.
In CFT, some of these symmetries were identified as different abelian
gaugings
of a WZW theory, \cite{K1}, and this was generalized to abelian
gaugings of arbitrary theories with chiral currents, \cite{RV}, and
organized into (semi-classical) O(d,d,Z) type
symmetries, \cite{GR}, mimicking the situation for flat backgrounds.
Moreover, for coset models, such duality symmetries exist also for
backgrounds without any isometries, \cite{K1,K2}.
Non-abelian duality received more attention recently,
\cite{nonabelian}
but
its status is not yet clear.
A careful analysis of the underlying CFT structure, revealed that
most of these semiclassical symmetries, pertaining to compact cosets,
are indeed exact in string theory, \cite{K2}, and they are intimately
related to
the affine Weyl symmetries of the ``parent" theory, the WZW model.

In the non-compact case, the affine Weyl group is not a manifest
symmetry
but it can be shown that a particular kind of duality, axial-vector
duality, \cite{K1} is still a
symmetry, \cite{GK}\footnote{
See also A. Giveon's talk in this volume.}.

At the semiclassical level, provided there is an abelian isometry,
the duality transformation can be effected by gauging this isometry
and adding also
a langrange multiplier coupled to the field strength of the gauge
field, \cite{buscher}.
Integrating out the langrange multiplier, forces the gauge field to
be pure gauge which can, subsequently, be gauged away, giving back
the original model.
On the other hand, one can gauge fix to a unitary gauge and then
integrate out the gauge field (which appears quadratically in the
action).
In this way, a different (dual) sigma model action is obtained (the
measure can be also taken care off, effectively changing the
dilaton).
Modulo global properties, \cite{AG}, the original and the dual action
describe
the same theory.

Duality has important implications for string theory.
It can be thought of as an unbroken part of the full string
symmetry in a particular background, and it can provide important
hints concerning the string physics around that background.
In particular, all possible dual backgrounds are relevant
for string propagation as each determines the response of some of the
string
modes.
It is also interesting that duality seems to preserves some general
relativistic notions as that of a Hawking temperature and entropy of
black holes, \cite{HW}.

Explicit studies in specific models, \cite{K2} (although the result
seems to be general)
indicate that, in the non-flat case, duality maps zero modes to
oscillator modes of the string. In some coordinate system, such
oscillator modes
resemble winding modes, the only difference being that they are not
``topological" (the target manifold has no non-contractible cycles).
I will comment more on the possible implications of duality for
string physics
in non-compact curved backgrounds in the last section.

\section{WZW Models}

In this section, we will analyze in detail the duality symmetries of
WZW models,
both from the $\s$-model and the CFT (affine current algebra) point
of view.

It turns out that understanding the simplest
group, SU(2), will suffice. In the case of non-simply laced simple
groups
there are some minor changes due to the short roots that will be
dealt with
latter on.
The case of non-simple groups has further complications that we will
not consider here.

The action of the WZW model is
$$I(g)={k\over 4\pi}I_{NS}(g)+{ik\over
6\pi}\Gamma_{WZ}(g)\eqno(2.1)$$
$$I_{NS}(g)=\int d^{2}x
Tr[U_{\mu}U^{\mu}]\;\;,\;\;\Gamma_{WZ}(g)=\int
\limits_{B\atop \p B=S^{2}}
d^{3}y\varepsilon^{\mu\nu\rho}Tr[U_{\mu}U_{\nu}U_{\rho}]\eqno(2.2)$$
where
$$U_{\mu}=g^{-1}\p_{\mu}g\;\;,\;\;V_{\mu}=\p_{\mu}g
g^{-1}\eqno(2.3)$$
$g$ is a matrix in the fundamental representation of $G$ and $Tr$ is
a properly normalized trace such that
$${1\over 12\pi^{2}}\int_{S^{3}}Tr[U\wedge U\wedge U]\in
Z\;.\eqno(2.4)$$
The action $I(g)$ is invariant under the group $G_{R}\otimes G_{L}$,
generated by left and right group transformations, $g\rightarrow
h_{1}gh_{2}$,
with associated conserved currents
$$J^{\mu}_{R}={k\over 2\pi}P_{-}^{\mu\nu}U_{\nu}\;\;,\;\;J^{\mu}_{L}=
{k\over 2\pi}P_{+}^{\mu\nu}V_{\nu}\eqno(2.5)$$
with $P_{\pm}^{\mu\nu}\equiv\delta^{\mu\nu}\pm
i\varepsilon^{\mu\nu}$.
These currents are conserved and chirally conserved and they generate
two copies of the affine $\hat G$ current algebra.
An important property of the WZW action is that it satisfies the
Polyakov-Wiegman formula
$$I(gh)=I(g)+I(h)-{k\over 2\pi}\int
d^{2}xP^{\mu\nu}_{+}Tr[U_{\mu}(g)V_{\nu}(h)]\eqno(2.6)$$

To generate duality transformations in the WZW model, we pick a
generator of the Lie algebra of $G$, $T^{0}$, normalized as
$Tr[(T^{0})^{2}]=1$.
We can then parametrize $g=e^{i\phi T^{0}}h$.
Using (1.6), the action $I(g)$ takes the form
$$I(g)=I(h)+{k\over 4\pi}\int \p_{\mu}\phi\p^{\mu}\phi -{ik\over
2\pi}
\int P^{\mu\nu}_{+}\p_{\mu}\phi V^{0}_{\nu}(h)\eqno(2.7)$$
where $V^{0}_{\mu}(h)=Tr[T^{0}V_{\mu}(h)]$.
We can now apply the duality map, \cite{K1} to
obtain\footnote{
The measure also changes by a finite computable piece, \cite{K1}}
$$I^{\rm dual}(g)=I(h)+{1\over 4\pi k}\int \p_{\mu}\phi\p^{\mu}\phi-
{i\over 2\pi}\int P_{+}^{\mu\nu}\p_{\mu}V^{0}_{\nu}(h)\;.\eqno(2.8)$$
The angle $\phi$ was originally normalized to take values in
$[0,2\pi]$.
It is obvious from (2.8) that the effect of the duality
transformation
is to change the range of values to $[0,2\pi/k]$.
To see how many independent duality transformations exist, we have to
explicitly parametrize the Cartan torus dependence of the WZW model.
Pick a basis in the Cartan algebra, $T^{i}$, $i=1,2,\cdots,r$,
$[T^{i},T^{j}]=0$, $Tr[T^{i}T^{j}]=\delta^{ij}$ and parametrize,
$$g=e^{i\sum_{i=1}^{r}\a^{i}T^{i}}\,h\,e^{i\sum_{i=1}^{r}\g^{i}T^{i}}
\;\;.\eqno(2.9)$$
Then using (2.6) the WZW action becomes
$$I(g)=I(h)+{k\over
4\pi}\int(\p_{\mu}\a^{i}\p^{\mu}\a^{i}+\p_{\mu}\g^{i}
\p^{\mu}\g^{i})-{ik\over
2\pi}\int(P_{+}^{\mu\nu}\p_{\mu}\a^{i}V^{i}_{\nu}
(h)+P_{-}^{\mu\nu}\p_{\mu}\g^{i}U_{\nu}^{i}(h))+$$
$$+{k\over 2\pi}\int
P_{+}^{\mu\nu}\p_{\mu}\a^{i}\p_{\nu}\g^{j}M^{ij}(h)\eqno(2.10)$$
where
$$U_{\mu}^{i}(h)=Tr[T^{i}U_{\mu}(h)]\;,\;V_{\mu}^{i}(h)=
Tr[T^{i}V_{\mu}(h)]
\;,\; M^{ij}(h)=Tr[T^{i}hT^{j}h^{-1}]\;.\eqno(2.11)$$
It is obvious from (2.10) that we can apply the duality
transformation
using  any of the $\a^{i}$, $\g^{i}$.
Thus, there are $2^{2r}-1$ non-trivial duality transformations.
A duality transformation on $\a^{i}$ effectively makes the
substitution
$\a^{i}\rightarrow \a^{i}/k$ in the action whereas a duality
transformation
on $\g^{i}$ makes the substitution $\g^{i}\rightarrow -\g^{i}/k$.
The new manifold has a similar metric to the group manifold but due
to the different angle periodicities, has Taub-NUT type
singularities.

In order to identify the underlying property of the WZW model,
responsible
for the invariance under these duality transformations, we have delve
a bit
into such elements of the the representation theory of the affine Lie
algebras
as the affine Weyl group and external automorphisms.

The affine Weyl group ${\hat W}$ is a semidirect product of the Lie
algebra
Weyl group $W$ times a translation group, ${\hat W}=W\triangleright
T$.
Apart from the action of finite Weyl group elements, there are Weyl
transformations associated to roots which have a component in the
direction of
the imaginary simple root.
The action of such an element ${\hat W}_{\vec\a}$
on a finite Lie algebra weight $\vec\lambda$ and on the grade $n$ is
$${\hat
W}_{\vec\a}({\vec\lambda})=W_{\vec\a}({\vec\lambda})-k{\vec\beta}
\eqno(2.12a)$$
$${\hat W}_{\vec\a}(n)=n-{\vec\lambda}\cdot{\vec\beta}-{k\over
2}{\vec\beta}\cdot{\vec\beta}\eqno(2.12b)$$
where ${\vec\beta}=2{\vec\a}/{\vec\a}\cdot{\vec\a}$ is the coroot
associated
to the finite Lie algebra root $\vec\a$, the grade $n$ is
basically the mode number\footnote{
In a highest weight representation where
the affine primaries have $L_{0}$ eigenvalue $\Delta$, the grade $n$
of a
state is the eigenvalue of $L_{0}-\Delta$ on that state.} and
$W_{\vec\a}
(\vec\lambda)={\vec\lambda}-{\vec\a}({\vec\lambda}\cdot{\vec\beta})$
is a
finite Weyl transformation.
It is important to note that affine Weyl transformations, in general,
map
states inside a representation at different levels.

There are also external automorphisms of the affine algebra which are
essentially associated to symmetries of the affine Dynkin diagram.
For the $SU(n)$ case, the affine Dynkin diagram consists of $n$ nodes
connected around a circle. The external automorphisms are generated
by a basic rotation, and a reflection which corresponds to the finite
Lie
algebra external automorphism (that maps a representation to its
complex
conjugate). When we write a highest weight
${\vec\Lambda}=\sum_{i=1}^{n-1}
m_{i}{\vec\Lambda}_{i}$ in terms of the fundamental weights
${\vec\Lambda}_{i}$, ($m_{i}$ are non-negative integers), the action
of the
generating rotation of the affine Dynkin diagram is as follows
$$\sigma({\vec\Lambda})=(k-\sum_{i=1}^{n-1}m_{i}){\vec\Lambda}_{1}+
m_{1}{\vec\Lambda}_{2}+\cdots+m_{n-2}{\vec\Lambda}_{n-1}\;.
\eqno(2.13)$$
$\sigma$ generates a $Z_{n}$ group\footnote{
In general this group is isomorphic
to the center of the finite Lie group} where $\sigma^{n}=1$ on the
highest weights, but acts as an affine Weyl transformation in the
representation.
Specializing to SU(2), let $m\in Z/2$ be the weight, and $j\in Z/2$
the highest weight (spin of a representation).
Then the finite Weyl group acts as $m\rightarrow -m$, and combined
with
the affine translation $m\rightarrow m+k$ they generate the affine
Weyl group.
The only nontrivial outer automorphism $\sigma$ acts as $j\rightarrow
k-j$
and $\sigma^{2}$ is a Weyl translation.

The non-trivial statement now is: For compact groups, integer level
and integrable highest weight representations, both the affine Weyl
group
and the external automorphisms are symmetries.
In particular, in a WZW model the Hilbert space is constructed by
tying
together (in a modular invariant way) two copies of representations
of the
affine algebra. Thus, we have invariance under independent affine
Weyl
transformations acting on left or right representations.
Moreover, since the modular transformation properties of the affine
characters
reflect the external automorphism symmetries, the theory is invariant
under external automorphisms that act at the same time on left and
right
representations.
These invariance properties can be verified for correlation functions
on the
sphere and the torus. This then implies that they hold on an
arbitrary Riemann
surface
since the sphere and torus data are sufficient in order to construct
the correlators at higher genus.

As an example, we will present the SU(2) case and focus on the
spectrum.
We introduce the (affine) SU(2)$_{k}$ characters
$$\chi_{l}(q=e^{2\pi i\tau}, w)=Tr_{l}\left[q^{L_{0}}e^{2\pi
iwJ^{3}_{0}}
\right]=\sum_{m=-k+1}^{k}
c^{l}_{m}(q)\vartheta_{m,k}(q,w)\eqno(2.14)$$
where $l$ is twice the spin (a non-negative integer) and  $m$ is
twice the $
J^{3}_{0}$ eigenvalue.
The trace is in the affine hw representation of spin $l$,
$$\vartheta_{m,k}(q,w)=\sum_{n\in Z} q^{k(n+{m\over 2k})^{2}}e^{2\pi
iw(kn+{m\over
2})}\eqno(2.15)$$
and $c^{l}_{m}$ are the standard string functions  which satisfy
$c^{l}_{m}=0$ when $l-m=1(mod$ $2)$ (which means that the spin is
increased
or decreased in units of 1) .
For integrable representations ($k$ is a positive integer and $0\leq
l\leq k$),
invariance under the affine Weyl group is equivalent to
$$c^{l}_{m}=c^{l}_{-m}\;\;,\;\; c^{l}_{m}=c^{l}_{m+2k}\eqno(2.16)$$
The first relation is due to the Weyl group of $SU(2)$ while the
second is the
generating translation in the affine Weyl group.
There is another important relation
$$c^{l}_{m}=c^{k-l}_{k-m}\eqno(2.17)$$
which is a consequence of the external affine automorphism.

The duality transformation on $\a^{i}$ amounts to replacing ${\bar
J}^{i}
\rightarrow -{\bar J}^{i}$, where ${\bar J}^{i}$ is the right Cartan
current
in the $T^{i}$ basis of the Cartan subalgebra.
Similarly the duality transformation on $\g^{i}$ amounts to the
replacement
$J^{i}\rightarrow -J^{i}$ at the level of the Cartan subalgebra.
This is not the whole story however. With a bit more effort one can
see that
they
act as Weyl transformations on the left or right SU(2) currents.
This identification can be seen clearly by coupling the WZW action to
external
gauge fields and monitoring the effect of the duality transformation
on the
currents. It can also be recovered  from the twisted partition
function
via the action of the duality transformation on the gauge field
moduli
(for the Cartan).

The duality transformations
$$D_{i}\;\;:\;\; J^{i}\rightarrow -J^{i}\eqno(2.18a)$$
$${\bar D}_{i}\;\;:\;\;{\bar J}^{i}\rightarrow -{\bar
J}^{i}\eqno(2.18b)$$
are exact symmetries of the model.
This can be verified explicitly, since characters are invariant under
the finite Weyl group.

This invariance is similar, but qualitatively different than that
present
in flat backgrounds. There, one has a family of theories parametrized
by $G,B$ and duality is the statement that two theories are
equivalent
for different values of the parameters. Here, there is no parameter
present
and, in this sense, this is what we could call self-duality.
It corresponds to the self duality (in the flat case) of level 1 WZW
models
appearing at special values of the moduli.
Now we are in a position to discuss the general WZW model for a
simple
group $G$. Let $M$ be the root lattice, $M_{L}$ the long root lattice
and $M^{*}$ the weight lattice.
The character of a hw representation of ${\hat G}$ with hw $\vec
\Lambda$
is defined as
$$\chi_{\vec \Lambda}(q,{\vec w})=Tr[q^{L_{0}}e^{2\pi i{\vec
w}\cdot{\vec
J}_{0}}]\eqno(2.19)$$
where ${\vec J}_{0}$ generates the cartan subalgebra of $G$.
The character admits the string function decomposition, \cite{kp},
$$\chi_{\vec \Lambda}=\sum_{{\vec \lambda}\in M^{*}/kM_{L}} c^{\vec
\Lambda}_{\vec \lambda}(q)\Theta_{\vec \lambda}({\vec
w},q)\eqno(2.20)$$
with $\Theta_{\vec\lambda}$ being the classical $\vartheta$-function
of level $k$ of the Lie algebra of $G$
$$\Theta_{\vec\lambda}({\vec w},q)=\sum_{{\vec\gamma}\in M_{L}}
q^{{k\over 2}\left({\vec\gamma}+{{\vec\lambda}\over k}\right)^{2}}
e^{2\pi i{\vec w}\cdot(k{\vec\gamma}+{\vec\lambda})}\;.\eqno(2.21)$$
The string functions are invariant under the Weyl group and Weyl
translations
$$c^{\vec\Lambda}_{w({\vec\lambda})}=c^{\vec\Lambda}_{\vec\lambda}
\;\;,\;\;
c^{\vec\Lambda}_{{\vec\lambda}+k{\vec\beta}}=
c^{\vec\Lambda}_{\vec\lambda}
\eqno(2.22)$$
where $w$ is a Weyl transformation and ${\vec\beta}\in M_{L}$.

The (left) generating duality transformations $D_{i}$ correspond to
Weyl
reflections generated by the simple roots ${\vec\a}_{i}$ which
implement
the transformations (2.18a).
The invariance of the spectrum (and partition function) is encoded in
the
fact, obvious from (2.21,22), that $\chi_{\vec\Lambda}$ is invariant
under
$w_{i}\rightarrow -w_{i}$.
Although $w_{{\vec\a}_{i}}$ do not commute, they do so when applied
to the
character, thus at the level of the partition function they generate
a group
isomorphic to $Z_{2}^{r}$.
However, at the level of correlation functions the (left) duality
group is
larger and in fact isomorphic to $W_{L}\times W_{R} /W_{D}$, where
$W_{L,R}$ are
the left(right) Weyl groups of the (finite) Lie algebra of the WZW
model
and $W_{D}$ is the diagonal Weyl group (whose action corresponds to
reparametrizations of the action).

We have seen already (from affine algebra representation theory)
that the (non-local) symmetry of the (compact and unitary) $g$-WZW
model
is larger:
It is generated by the left and right affine Weyl groups $\hat
W^{g}_{L,R}$as well as the external affine automorphisms $A^{g}$.
I conjecture that $\hat W^{g}_{L}\times \hat W^{g}_{R}\times
A^{g}/W_{D}$ is the full self-duality group of the WZW.
Afine Weyl transformations act semi-classically as GL(r,Z) rotations,
that is, as particular changes of basis in the lattice of weights.
This implies, that although semi-classically, for curved backgrounds
$GL(r,Z)$ is a symmetry, \cite{GR}, in the exact theory only some
part
of it survives.
Concerning the external affine automorphisms, their action in
$\s$-model
language is not known.

\section{Compact Coset Models}

{}From the WZW theory, we can built other CFTs by projections. The
simplest such projection corresponds to constraining
the affine currents of a subalgebra, known as coset
construction, \cite{BH}.\footnote{
There are more general projections though, preserving conformal
invariance, \cite{HK}.}
The $\s$-model action of coset models is obtained by gauging the
appropriate
subgroup of the WZW model.
Gauging different dual versions of the WZW model, dual versions of
the coset model are obtained.
It can be shown that the Killing symmetries of a coset model $G/H$
are of two types, \cite{K2}:
Chiral $H'_{L}\times H'_{R}$ isometries, when there is a subgroup
$H'$ such that $[H,H']=0$ and non-chiral abelian isometries in one
 to one correspondence with U(1) factors of $H$.
In many cases, the aforementioned  duality exists for
coset actions without isometries, \cite{K2}.

A special form of duality is obtained for coset models where the
gauged subgroup contains a U(1) factor.
In such a case, one has the option of gauging either the axial or the
vector
subgroup of the original $U(1)_{L}\times U(1)_{R}$ subalgebra.
The $\s$-model actions of these two gauged models are generically
different
but it can be shown that the models are dual to each other,
\cite{K1}.
This type of duality is known as axial-vector duality and at the
semiclassical
level is powerful in generating different types of
backgrounds, \cite{RV,GR}.
It would seem that Weyl symmetry of the affine algebra is enough to
guarantee
that axial-vector duality is an exact symmetry.
The story is more complicated though. It turns out, \cite{K2} that
the
underlying symmetry of current algebra responsible for axial-vector
duality
is affine-Weyl symmetry, $\hat W_{L}\times \hat W_{R}$.
For integrable (unitary) representations of (compact) affine algebras
the affine Weyl group is a symmetry and so is axial-vector duality.

The full duality symmetry of a coset model $g/h$ is generated by the
full duality symmetry of the ``parent" $g$-WZW theory.
Let $H$ be a reductive subgroup of $G$ and $H^{na}$ be its
non-abelian
component, (discard U(1) factors). Denote the Lie algebra of $H^{na}$
by
$h$.
Then, $D^{h}=\hat W^{h}_{L}\times \hat W^{h}_{R}\times A^{h}$ is a
normal
subgroup of $D^{g}=\hat W^{g}_{L}\times \hat W^{g}_{R}\times A^{g}$.
The full self-duality group of $G/H$ is $D^{g}/D^{h}$.
The need to factor $D^{h}$ comes since it acts only on the gauge
degrees
of freedom and it is thus, invisible in the $G/H$ theory.

\section{Duality Symmetries in the Moduli Space of WZW
and Coset Models.}

So far we have been concerned with the self duality symmetries of WZW
and
coset models. We will now show how this self duality generates
duality symmetries of their moduli space.

The primary example of this is the moduli space of (flat)
D-dimensional toroidal models. At special points of the moduli, one
finds
WZW models (at level one). It can be shown in this case that, the
full duality group of the moduli space, $O(D,D,Z)$ can be generated
from the
self duality symmetries of the WZW points, \cite{GMR}.

A similar phenomenon happens in the general case.
Consider a WZW (or coset) model and its neighborhood in moduli space
which
is generated via marginal perturbations by integrable (1,1) operators
$O^{i}_{1,1}$.
This set of operators contains dual pairs, that is for each
$O^{i}_{1,1}$
there is a $\tilde O^{i}_{1,1}$ in the set related by a self-duality
of the WZW or coset theory.
At the level of the (abstract) conformal field theory, $O^{i}$ and
$\tilde O^{i}$ correspond to the same operator.
In a $\s$-model (field theoretic) realization, they are different
fields.

Self-duality at the WZW or coset point implies that the line
generated
by $O^{i}$ is equivalent as a CFT to that generated by $\tilde
O^{i}$.
The backgrounds ($\s$-models) corresponding to the two lines though
are different.
Thus, in the full moduli space, duality symmetries can be
generated
by the self-duality symmetries at special points (WZW and
cosets)

In the case of a G-WZW model for simple G, at a generic level the
only integrable perturbations
are generated by $J^{i}\bar J^{j}$ constructed out of the left and
right Cartan
currents.
Thus (except at special points) the dimension of the moduli space is
$D^2$ where $D=rank_{G}$.
The duality group here is similar to O(D,D,Z), since in such
deformations
the generalized G-parafermion theory does not change along the moduli
space
and duality acts on the cartan-torus bosons.
When G is semi-simple, then one has extra generic (1,1) operators,
however their integrability is an open question.

Similar remarks apply to perturbations by relevant operators. In this
case
self-duality of the WZW or coset theory implies Krammers-Wanier
duality
for the off-critical theory.
The archetype of this duality exists in the $Z_{N}$ parafermion
models perturbed by the first energy operator, which upon a
self-duality
transformation in the fixed point theory, changes sign.

It is tempting to conjecture that all duality symmetries of the full
moduli
space of WZW and coset models are reflections of self-duality at
these special
points, as it happens in the flat case.

\section{Duality in the Non-compact Case.}

So far we have dealt with compact WZW models and their cosets.
In the non-compact case things can, a priori, be different.
At the level of duality symmetries of the G-WZW model itself, those
associated
with the finite Weyl group are still symmetries (if G is reductive).
However the affine Weyl acts differently.
For a generic affine representation, affine Weyl transformations map
it
to different representations.
We have seen that affine Weyl symmetry is important for axial-vector
duality.
The only way that axial-vector duality can survive in the non-compact
case
is, if the spectrum is organized into complete orbits of the affine
Weyl group.

There is a different way of showing that axial-vector duality remains
an
exact symmetry in the non-compact case, \cite{GK}.
This, in retrospect implies the organization of the spectrum as
mentioned above.
Important information about the spectrum can be retrieved that way
since, (at least for the case of SL(2,R)) non-compact string
functions
are known, \cite{BK} as well as their transformation properties under
the affine
Weyl group.
It is an interesting problem to try to obtain the spectrum this way.

\section{Comments on the Physics of Duality.}

The main lesson from duality and related symmetries is that the
background
fields do not determine uniquely the spectrum and physics of string
theory.
Duality can be viewed as a tiny (unbroken) part of the huge string
gauge symmetry, whose glory remains obscure to our days.
Another way to state this, is, that different modes of the string
feel different geometry.
Thus, the background interpretation of string vacua should be used
with care
in order to ascertain the physics.
The only cases were the description is reliable is the large volume
limit
of compact manifolds, and at the asymptotically flat region of
non-compact manifolds. Once at a region of finite curvature, the
geometrical
description of string theory breaks down.
Even topology is not preserved under duality and
there are continuous families
of ground states in string theory where topology changes without the
occurrence of anything catastrophic, \cite{W1,AGM,GK}.

An example of how this type of symmetry can affect string
propagation, can be given (heuristically) as
follows\footnote{
This argument is advanced in collaboration with C. Kounnas.}.
Consider a string background which is highly curved or even singular
(semi-classically) in a certain region, (the 2-d black hole,
\cite{W},
is such an example but one needs to add a few extra dimensions in
order to have non-trivial massive states).
In the asymptotic region, (which is obtained by some
spacetime-depended radius becoming very large), one has quantum
numbers for asymptotic states that
correspond roughly to windings and momenta. Momentum states are the
only low energy states in this region.
Consider a momentum mode travelling towards the high curvature
region.
Its effective mass starts growing as it approaches large curvatures.
At some point it becomes energetically possible for it to decay to
winding
states which, in this region, start having effective masses that are
lower
than momentum modes.
In such backgrounds (unlike flat ones) winding and momentum are not
separately conserved so that such a transition is possible.
The reason for this is that there is a non-trivial dilaton field and
thus, winding and momentum conservation is broken by the screening
operators
which transfer it to discrete states localized at the high curvature
region. An alternative interpretation of this, is that particles
interact
with such localized states loosing momentum (in discrete steps) and
gaining
winding number.

Once such a momentum to winding mode transition happens in the
strongly curved
region, the winding state sees a different geometry, namely the dual
one and thus continues to propagate further into the strong curvature
region
since it feels only the (weak) dual curvature.
This phenomenon, implies the need of a novel treatment of the
underlying effective field theory approach, which would resemble the
effective treatment
of particles in condensed matter physics, with position dependent
masses.

This type of picture indicates that the physics of string black holes
will be qualitatively different that their classical general
relativity counterparts.

\vfill\eject

\end{document}